\documentclass[aps,prb,a4paper,preprint,showpacs]{revtex4}
\usepackage{amsmath}
\usepackage{amsfonts}
\usepackage{amssymb}
\usepackage{slashbox}
\usepackage{graphicx}
\usepackage{amsbsy}

\begin{document}
\title{Spin-dependent Fano resonance induced by conducting chiral helimagnet contained in a quasi-one-dimensional electron waveguide }
\author{Rui Zhu\renewcommand{\thefootnote}{*}\footnote{Corresponding author. Electronic address:
rzhu@scut.edu.cn}}
\address{Department of Physics, South China University of Technology,
Guangzhou 510641, People's Republic of China }

\begin{abstract}

Fano resonance appears for conduction through an electron waveguide
containing donor impurities. In this work, we consider the thin-film conducting chiral
helimagnet (CCH) as the donor impurity in a one-dimensional waveguide model. Transmission and conductance for arbitrary CCH spiral period are obtained. Due to the spin spiral coupling, interference between the direct and intersubband transmission channels gives rise to spin-dependent Fano resonance effect. The spin-dependent Fano resonance is sensitively dependent on the helicity of the spiral. By tuning the CCH potential well depth and the incident energy, this provides a potential way to detect the spin spiral period in the CCH.

\end{abstract}


\maketitle

\narrowtext

\section{Introduction}

Fano resonance, characterized by resonant reflection results from constructive and destructive interference of two quantum paths across localized and extended states\cite{Ref1, Ref2, Ref8}. It is a universal effect existing almost in all interfering quantum processes independent of the specific details of the system under study. The interfering paths can be formed by spatial inhomogeneity as well as time-dependent oscillation. Subsequent to its discovery, there have been a great number of studies devoted to Fano resonances in various
quantum systems, such as Anderson impurity systems\cite{Ref9}, quantum dots\cite{Ref10, Ref6, Ref23}, scattering from a donor impurity in an electron waveguide,\cite{Ref2, Ref3}, tunneling through an ${\rm{Al}}_{\rm{x}} {\rm{Ga}}_{{\rm{1 - x}}} {\rm{As}}$ barrier\cite{Ref2, Ref4}, transmission through a
waveguide linked to a resonant cavity\cite{Ref2, Ref5}, spin inversion devices\cite{Ref24}, nanowires and tunnel
junctions\cite{Ref11} and etc.\cite{Ref8}. Recently, the Fano
resonance has been found\cite{Ref7} in plasmonic nanoparticles, photonic crystals, and electro magnetic metamaterials. The steep dispersion
of the Fano resonance profile promises applications in sensors, lasing, switching, and nonlinear and slow-light devices\cite{Ref7}.

Chiral helimagnets are static magnetic states sustained in various materials lacking rotoinversion symmetry with Dzyaloshinkii-Moriya (DM) antisymmetric exchange
interaction\cite{Ref13}. It is also the ground magnetic structure of most multiferroic materials. Recently, the spatial-dependent spin structure were exploited in different functional devices such as persistent spin currents\cite{Ref19}, spin-field-effect transistor\cite{Ref17}, tunneling anisotropic magnetoresistance\cite{Ref18}, spin resonance\cite{Ref13} and spin diffraction\cite{Ref14, Ref15, Ref16}. It has been shown that different transverse tunnels of transmission are coupled in free electron and spiral helimagnet hybrid structures\cite{Ref14, Ref15, Ref16}. A possible experimental
candidate of a conducting chiral helimagnet (CCH)\cite{Ref12},
where free electrons are
coupled with the background helimagnetic texture via the sd-type
interaction is currently available in\cite{Ref12}
${\rm{Cr}}_{{{\rm{1}} \mathord{\left/
 {\vphantom {{\rm{1}} {\rm{3}}}} \right.
 \kern-\nulldelimiterspace} {\rm{3}}}} {\rm{NbS}}_{\rm{2}} $. The sd-exchange interaction gives rise to a spatial dependent Zeeman energy. We consider a quasi-one-dimensional wave guide containing a thin CCH well and investigate the transport properties in it. When the CCH film acts as a donor impurity with a space-dependent potential in the spin space, spin-dependent Fano resonance occurs. In the electron waveguide configuration, Fano conductance is sensitively dependent on the CCH helicity, which provides a potential measurement of the spin spiral period in the CCH.

\section{Theoretical formulation}

The device we propose is a quasi-1D electron waveguide embedded with a thin layer of CCH acting as a donor impurity, which is depicted in Fig. 1. The Schr\"{o}dinger equation describing scattering in such a device is
\begin{equation}
\left[ { - \frac{{{\hbar ^2}}}{{2m}}{\nabla ^2} + {V_c}\left( x \right) + {V_{sc}}\left( {x,z} \right)} \right]\psi \left( {x,z} \right) = E\psi \left( {x,z} \right).
\end{equation}
Confining potential $V_c \left( x \right)$
of the wave guide is an infinite square well across the $x$-direction, with eigenstates
\begin{equation}
\phi _n \left( x \right) = \sqrt {2/a} \sin \left( {n\pi x/a} \right),
\end{equation}
and corresponding subband energies
\begin{equation}
\varepsilon _n  = \hbar ^2 \pi ^2 n^2 /2ma^2
\end{equation}
 with $n$ the subband index and $m$ the free-electron mass. We consider an ultrathin CCH layer, that is, the localized spin-dependent scattering potential can be approximated by a Dirac-delta
function as
\begin{equation}
V_{sc} \left( {x,z} \right) = \left( {\tilde J{\bf{n}}_r  \cdot {\bf{\sigma }} - V_0 d} \right)\delta \left( z \right),
\end{equation}
where $\tilde J$ is the sd-type exchange coupling
strength between free electrons and the background spin texture, ${\bf{n}}_{\bf{r}}  = \left[ {\sin \bar qx,0,\cos \bar qx} \right]$ is the local magnetization direction with $\bar q =2 \pi / \lambda $ the spin wave vector of the spiral ($\lambda $ is the spiral period), $\bf{\sigma }$ is the Pauli vector, and $V_0$ and $d$ are depth and width of the potential well, respectively\cite{Ref25}. Arbitrary spiral periods $\lambda$ relative to the waveguide width $a$ can be considered as $\lambda  = {{\left( {fa} \right)} \mathord{\left/
 {\vphantom {{\left( {fa} \right)} l}} \right.
 \kern-\nulldelimiterspace} l}$ with $f$ and $l$ arbitrary positive integers.

Using the waveguide eigenstates $\left\{ {\phi _n } \right\}$ as the expanding basis of the wave function $\psi \left( {x,z} \right)$, Eq. (1) can be converted to a set of linear
equations. In the matrix formulism, the equations are\cite{Ref2, Ref20}
\begin{equation}
\left[ { - 2iK_{mn}^{\alpha \beta }  + V_{mn}^{\alpha \beta } } \right]T_{nl}^{\beta \gamma }  =  - 2iK_{ml}^{\alpha \gamma } ,
\end{equation}
where $
K_{mn}^{\alpha \beta }  = k_m \delta _{mn}  \otimes I_{\alpha \beta }
$ is elements of the diagonal matrix of wave vectors with $ k_m=
\sqrt {2m\left( {E - {\varepsilon _m}} \right)} /\hbar
$ and is extended to the wave-spin product space.
Also, elements of the spin-dependent interband transition
matrix
\begin{equation}
V_{mn}^{\alpha \beta } = \frac{{2m}}{{{\hbar ^2}}}\left\langle {\phi _m^\alpha } \right|\tilde J{{\bf{n}}_r} \cdot {\bf{\sigma }} + {V_0}d\left| {\phi _n^\beta } \right\rangle ,
\end{equation}
where
\begin{equation}
\left| {\phi _m^ u  } \right\rangle  = \psi _m  \otimes \left( {\begin{array}{*{20}c}
   1  \\
   0  \\
\end{array}} \right),\begin{array}{*{20}c}
   {} & {\left| {\phi _m^ d  } \right\rangle  = \psi _m  \otimes \left( {\begin{array}{*{20}c}
   0  \\
   1  \\
\end{array}} \right),}  \\
\end{array}
\end{equation}
expressed in the $\sigma _y$ representation with spin ${\left( {\begin{array}{*{20}{c}}
1&0
\end{array}} \right)^T}$ state parallel to the $y$-coordinate and vice versa. Expressions of $V_{mn}^{\alpha \beta }$ are provided in the Appendix for reference. When the embedded CCH spiral shifts the sinusoidal phase by $\phi $ as ${\bf{n}}_{\bf{r}}  = \left[ {\sin (\bar qx+\phi),0,\cos (\bar qx+\phi)} \right]$, $V_{mn}^{\alpha \beta }$ can be analogously obtained. Similarly, the procedure can be extended to spin spirals spanned in different space planes and three-dimension spirals.

In Eq. (5) Einstein notation is employed with the upper and lower footnotes indexing the spin and subband elements respectively.
Therefore, the spin-dependent transmission amplitudes $T_{nl}^{\beta \gamma }$ can be obtained from Eq. (5) by matrix algebra. Here $T_{nl}^{\beta \gamma }$ denotes transmission amplitudes from the $l$-th subband with spin-$\gamma$ polarization to the $n$-th subband with spin-$\beta$ polarization. Accordingly, the reflection amplitudes can be obtained by
\begin{equation}
{\bf{R}} = {\left\{ {{{\left[ {2i{\bf{K}} - {\bf{V}}} \right]}^{ - 1}}{\bf{V}}} \right\}^{Trans}}.
\end{equation}
Indexes of matrixes $\bf{K}$, $\bf{V}$, and $\bf{T}$ are defined in Eq. (5). Elements of $\bf{R}$ are defined following that of $\bf{T}$. Footnote ``$Trans$" indicates the transpose of the matrix.
The spin-dependent conductance can be calculated from the two-terminal Landauer formula\cite{Ref2},
\begin{equation}
{G^ \alpha } = \frac{{{e^2}}}{h}\sum\limits_{mn\beta } {\left( {\frac{{{k_m}}}{{{k_n}}}} \right)T_{mn}^{*\alpha \beta }T_{mn}^{\alpha \beta }} ,
\end{equation}
where $\alpha$ indexes the spin-up or down channel and
the $m$ and $n$ are summed only over the propagating
modes.

The above derivation used the $\sigma _y$-representation, which gives rise to spin-up and down symmetry in all the transmissions, i.e. $T^{u u}=T^{d d}$, $T^{u d}=T^{d u}$, and thus $G^{u} =G^{d}$. If we consider ferromagnetic leads polarized in arbitrary directions, that symmetry would not survive. Take the $\sigma _z$-representation as the simplest alternative, we have
\begin{equation}
{T^{{\sigma _z}}} = U{T^{{\sigma _y}}}{U^{ - 1}},
\end{equation}
with
\begin{equation}
U = \frac{1}{{\sqrt 2 }}\left( {\begin{array}{*{20}{c}}
1&1\\
1&{ - 1}
\end{array}} \right),
\end{equation}
transforming the spin space. Similarly, results of arbitrary ferromagnetic polarization can be obtained using different $U$-matrixes.
Spin channels can thus be distinguished in the conductance.

\section{Numerical results and interpretations}

We consider the transmission properties of the CCH-layer-embedded quasi-1D electron waveguide. In numerical calculations, the sd-exchange coupling strength $\tilde J = 61.725$
${\rm{meV}} \cdot {\rm{nm}}$, which is reasonable compared to other
energy scales in the considered system. In our model, the width of the waveguide $a=5$ nm and the CCH spiral period $\lambda$ can vary continuously from $a/2$ to $18a$ with the spiral vector $\bar q= 2\pi / \lambda$. Namely, $\lambda =2.5$ nm to $90$ nm, which spans from short-period spirals ($3 \sim 6 $ nm) to long-period spirals ($18 \sim 90$ nm), which are typical values from experimental observation\cite{Ref21}. Potential well width of the CCH plane
$d=2$ nm and the potential well depth can be continuously tuned by gate voltages applied on the CCH layer\cite{Ref25}.

We set the incident energy $E$ between the first and second subbands of the waveguide with the eigen-energies $\varepsilon _1  \approx 15$
and $\varepsilon _2  \approx 60$ meV. When the lateral width of the waveguide is small giving rise to large energy spacing between subbands, the two-subband cutoff is an acceptable approximation. Thus we neglect all but the first and second subbands and solve the resulting
$4 \times 4$ matrix equation. So that the
only incident wave is in the lowest subband with either spin component.

When a spiral helimagnet layer, conducting or insulating, is contained in a 2D waveguide or in free space transport, spin-dependent diffraction can be seen\cite{Ref16}, i.e., spin-spiral wave vectors are added to or subtracted from the incident momentum wave vector and the electron is diffracted to different directions with spin reoriented. However, in a quasi-1D waveguide, diffraction is not sustained due to mode connection on the two sides of the CCH layer. In this situation, spin-dependent interband transition produces interference between direct and intersubband transmission channels resulting in the Fano resonance.

To demonstrate the Fano residence, we give numerical results of the transmission and conductance for skipped values of the CCH spiral period $\lambda =a/2$, $a$, $2a$, $4a$, and $6a$ in Figs. 2 to 5. As the inter-subband scattering potential acts in the spin space, difference between the spin-conserved and spin-flipped transmission is prominent. Spin polarization of the incident electron depends on the reservoirs at the ends of the waveguide. In our approach, $\sigma _y$-representation is used, which gives rise to spin-up and down symmetry in all the transmissions, i.e. $T^{u u}=T^{d d}$ and $T^{u d}=T^{d u}$. This symmetry would be modified when the incident spin polarization is not (anti)parallel to the $y$-direction.

It is known that the transmission probability of a waveguide embedded with a plane $\delta$-potential well or barrier is ${{4k_1^2} \mathord{\left/
 {\vphantom {{4k_1^2} {\left[ {4k_1^2 + {{\left( {{{2m{V_0}d} \mathord{\left/
 {\vphantom {{2m{V_0}d} {{\hbar ^2}}}} \right.
 \kern-\nulldelimiterspace} {{\hbar ^2}}}} \right)}^2}} \right]}}} \right.
 \kern-\nulldelimiterspace} {\left[ {4k_1^2 + {{\left( {{{2m{V_0}d} \mathord{\left/
 {\vphantom {{2m{V_0}d} {{\hbar ^2}}}} \right.
 \kern-\nulldelimiterspace} {{\hbar ^2}}}} \right)}^2}} \right]}}$
without resonance or antiresonance. With spatially-modulated $\delta$-potential scattering, the lowest evanescent mode is correlated with the propagating mode. Interference between the direct tunneling path and that through inter-subband scattering gives rise to resonant or antiresonant transmission. In our considered structure, the scattering potential acts in the spin space. Both the spin-conserved and flipped transmission demonstrates resonance or antiresonance at certain incident energy and CCH-well depth. Diagonal elements of the transition matrix $V_{mn}^{\alpha \beta }$ are $2m V_0 d / \hbar ^2$ while the cross elements depends on the CCH spiral structure. The potential well depth and the interband scattering strength coact to the transmission probability. For all the spiral structure considered, resonance or Fano resonance can be found at certain $V_0$ and $E$ parameter set up.

Figs. 2 to 5 are transmission results for different spiral periods. For $\lambda = 2a$, $4a$, and $6a$, only $1/2$, $1/4$, and $1/6$ of the spiral period are within the waveguide, respectively. Geometry of the scattering potential "seen" by the transporting electron includes different waveguide modes. Complete reflection occurs for both spin-conserved and flipped transmission, which indicates a Fano resonance. As the spiral period increases difference between different $\lambda$ becomes less prominent.

$\lambda = a$ is a special case as the spiral period equals the waveguide width. The $x$-component of the magnetization spirals sinusoidally and the $z$-component of the magnetization spirals consinusoidally in space. The sinusoidal component of the spin spiral $\sin \bar qx$ coincides with the second waveguide mode $
\phi _2 \left( x \right) = \sqrt {2/a} \sin \left( {2\pi x/a} \right)
$. As a result, spin-flipped transmission is dramatically suppressed demonstrating approximate complete reflection in all parameter regimes. And spin-conserved transmission demonstrates slow-varying antiresonance without considerable resonance. This can be explained by strong path interference of the tunneling process.

The case is different for $\lambda =a/2$. There are two spiral periods within the waveguide. The spiral variation $\sin \bar qx$ coincides with higher waveguide mode $
\phi _4 \left( x \right) = \sqrt {2/a} \sin \left( {4\pi x/a} \right)
$. The eigenenergy of this mode $\varepsilon _4  \approx 241$ meV is far away from the first and second mode energy, thus contributes little to interband transition and are neglected. Spin-conserved transmission demonstrates an antiresonance and spin-flipped transmission demonstrates a resonance.

For all the spiral periods, resonance peak in spin-conserved transmission occurs exactly at the energy of the antiresonance valley in spin-flipped transmission and vice versa. This can be interpreted by the scatterer profile. The sinusoidally-space-dependent spin exchange coupling coacts with the $\delta $-potential-well. Without the exchange coupling, transmission through a $\delta $-well in a waveguide varies monotonously in the $E$-$V_0$ parameter plane. The spin exchange coupling modulates this transmission probability in the spin space. Namely, the plane $\delta $-well transmission is enveloped by that modulation in spin space. The total transmission including both the spin-conserved and spin-flipped one is that of a plane $\delta $-well. That beam of spin-up incidence is split with spin-up wave transmitted and spin-down wave reflected. In this way, the CCH spiral also works as a spin splitter with fixed spiral period and potential well depth.

Numerical results of the conductance are given in Fig. 6. Due to $\sigma _y$ symmetry of the spiral and our representation setup, conductance in the spin-up channel $G^{u}$ equals that in the spin-down channel $G^{d}$. Since ${T^{uu}} > {T^{ud}}$ and ${T^{dd}} > {T^{du}}$, spin-conserved transmission dominates the conductance. This is an effect of spiral chirality of the CCH. There would be one spin dominance for a particular spin handedness or chirality analogous to circularly-polarized light traversing a spiral grating\cite{Ref22}. We used right-handed spiral CCH throughout our calculation, which gives rise to the spin-conserved transmission dominance.
It can be seen that the conductance sensitively depends on the spiral period of the CCH, which may lend a potential way to measure the magnetization configuration of certain materials. To further demonstrate this point we give numerical results of the conductance for continuously varying CCH spiral periods in Fig. 7.

It can be seen in Fig. 7 that for $\lambda $ smaller than $a/2$, spin-conserved transmission dominates spin-flipped transmission. As multiple spiral periods are embedded in the waveguide, spin-dependent transmission diffraction transits into interference. Spin-up (in the $\sigma _y$ representation) transmission is dramatically enhanced as right-handed spiral chirality is considered. For $\lambda $ larger than $a$, spin-up transmission is larger than spin-down one as well also due to the spiral chirality. As the spiral period $\lambda $ is increased to multiple $a$, only a small segment of the spatial variation is included in the waveguide. Geometry of the scattering potential "seen" by the transporting electron includes different waveguide modes. As the spiral period increases, difference between different $\lambda$ becomes less prominent. As discussed above, $\lambda =a$ and $a/2$ are two special cases with prominent Fano resonance. It is worth noting that the transmission and hence the conductance change sharply at particular spiral periods $\lambda =a/2$ and $a$, which suggests robust potential measurement of the CCH spiral period using the 1D waveguide Fano resonance effect.

We have also calculated conductance for the $5$ nm-wide waveguide with CCH-film embedded at other $V_0$ and $E$ values, and found that sharp increase or decrease is a general phenomenon and occurs at different small $\lambda$. Also, when the embedded CCH spiral shifts the sinusoidal phase by $\phi $ as ${\bf{n}}_{\bf{r}}  = \left[ {\sin (\bar qx+\phi),0,\cos (\bar qx+\phi)} \right]$, similar Fano effect can be expected. Since the sensitivity occurs at small $\lambda$ relative to $a$, wider waveguides should be used for long-period spirals. It can be seen from Eq. (3) that the waveguide eigen-energy spacing increases in square of $n$. Therefore, for both small and large waveguide widths, the two-subband cutoff is an acceptable approximation with the third subband level far above the second one. Numerically, we considered waveguides' width up to $50$ nm, prominent Fano effect and sharp conductance jump or fall occur as well. In our theoretical consideration, we do not go to details of experimental setup. We provide analytical formulation with numerical solutions referable for experimentalists.

In above numerical results, $\sigma _y$ representation is used. In this representation, spin-up conductance is equal to spin-down conductance. When we use ferromagnetic leads with spin polarized in the $y$ direction, no spin filtering effect occurs. However, for arbitrarily-polarized ferromagnetic leads, spin filtering effect is measurable. Fig. 8 shows numerical results of spin-up and down conductance in the $\sigma _z$ representation. The two spin channels are distinguished with prominent Fano resonance eminent in both channels.

\section{Conclusions}

Spin-dependent Fano resonance in a CCH-film-embedded quasi-1D electron waveguide was investigated. The CCH layer acts as a donor impurity with sinusoidally-space-dependent spin-exchange coupling. From numerical results of the transmission and conductance for arbitrary CCH spiral periods $\lambda$, it was found that the transport probabilities differ conspicuously for different relation between $\lambda$ and the waveguide width $a$. It was also found that by tuning the CCH potential well depth, transmission varies with spin-conserved and spin-flipped resonant or anti-resonant tunneling occurring at different incident electron energy. The proposed device may have potential application in spintronics and spiral magnetization determination.

\section{Acknowledgements}

The author acknowledges enlightening discussions with Zhi-Lin Hou, Wen-Ji Deng, and Jamal Berakdar. This project was supported by the National Natural Science
Foundation of China (No. 11004063) and the Fundamental Research
Funds for the Central Universities, SCUT (No. 2012ZZ0076).

\section{Appendix}

The spin-dependent interband transition matrix $V_{mn}^{\alpha \beta }$ for arbitrary $f$ and $l$ with $\lambda  = {{\left( {fa} \right)} \mathord{\left/
 {\vphantom {{\left( {fa} \right)} l}} \right.
 \kern-\nulldelimiterspace} l}$ can be derived as follows. Using the $\sigma _y$ representation, matrix scattering potential in the spin space
\begin{equation}
\tilde J{{\bf{n}}_r} \cdot {\bf{\sigma }} + {V_0}d = \left( {\begin{array}{*{20}{c}}
   {{V_0}d} & {\tilde J\left( {\cos \bar qx - i\sin \bar qx} \right)}  \\
   {\tilde J\left( {\cos \bar qx + i\sin \bar qx} \right)} & {{V_0}d}  \\
\end{array}} \right),
\end{equation}
which is hermitian.
\begin{equation}
\begin{array}{l}
 V_{nm}^{ u  u } = \frac{{4m}}{{a{\hbar ^2}}}\int {\left[ {\sin \left( {\frac{f}{{2l}}n\bar qx} \right)\left( {\begin{array}{*{20}{c}}
   1 & 0  \\
\end{array}} \right)\left( {\tilde J{{\bf{n}}_r} \cdot {\bf{\sigma }} + {V_0}d} \right)\sin \left( {\frac{f}{{2l}}m\bar qx} \right)\left( {\begin{array}{*{20}{c}}
   1  \\
   0  \\
\end{array}} \right)} \right]} dx \\
  =\frac{{4m}}{{a{\hbar ^2}}} \int {\sin \left( {\frac{f}{{2l}}n\bar qx} \right)} {V_0}d\sin \left( {\frac{f}{{2l}}m\bar qx} \right)dx =\frac{{2m{V_0}d}}{{{\hbar ^2}}}{\delta _{n,m}}. \\
 \end{array}
\end{equation}
For integer ${{2l} \mathord{\left/
 {\vphantom {{2l} f}} \right.
 \kern-\nulldelimiterspace} f}$,
\begin{equation}
\begin{array}{l}
V_{nm}^{ u  d } = \frac{{4m}}{{a{\hbar ^2}}}\int {\left[ {\sin \left( {\frac{f}{{2l}}n\bar qx} \right)\left( {\begin{array}{*{20}{c}}
1&0
\end{array}} \right)\left( {\tilde J{{\bf{n}}_r} \cdot {\bf{\sigma }} + {V_0}d} \right)\sin \left( {\frac{f}{{2l}}m\bar qx} \right)\left( {\begin{array}{*{20}{c}}
0\\
1
\end{array}} \right)} \right]} dx\\
 = \frac{{4m}}{{a{\hbar ^2}}}\int {\sin \left( {\frac{f}{{2l}}n\bar qx} \right)} \tilde J\left( {\cos \bar qx - i\sin \bar qx} \right)\sin \left( {\frac{f}{{2l}}m\bar qx} \right)dx\\
 = \frac{{m\tilde J}}{{{\hbar ^2}}}\left[ {\delta \left( {n,m + \frac{{2l}}{f}} \right) + \delta \left( {n + \frac{{2l}}{f},m} \right) - \delta \left( {n + m,\frac{{2l}}{f}} \right)} \right.\\
\left. { + i\frac{{{{\left( { - 1} \right)}^{n + m - \frac{{2l}}{f}}} - 1}}{{\pi \left( {n + m - \frac{{2l}}{f}} \right)}} + i\frac{{{{\left( { - 1} \right)}^{n - m + \frac{{2l}}{f}}} - 1}}{{\pi \left( {n - m + \frac{{2l}}{f}} \right)}} - i\frac{{{{\left( { - 1} \right)}^{n + m + \frac{{2l}}{f}}} - 1}}{{\pi \left( {n + m + \frac{{2l}}{f}} \right)}} - i\frac{{{{\left( { - 1} \right)}^{n - m - \frac{{2l}}{f}}} - 1}}{{\pi \left( {n - m - \frac{{2l}}{f}} \right)}}} \right].
\end{array}
\end{equation}
In Eq. (14), the term is zero when its denominator is zero.
For non-integer ${{2l} \mathord{\left/
 {\vphantom {{2l} f}} \right.
 \kern-\nulldelimiterspace} f}$,
\begin{equation}
\begin{array}{l}
V_{nm}^{ u  d } = \frac{{m\tilde J}}{{{\hbar ^2}}}\left[ {\frac{{{{\left( { - 1} \right)}^{n - m + 1}}\sin \left( {\frac{{2l\pi }}{f}} \right)}}{{\left( {n - m - \frac{{2l}}{f}} \right)\pi }} - \frac{{{{\left( { - 1} \right)}^{n + m}}\sin \left( {\frac{{2l\pi }}{f}} \right)}}{{\left( {n + m + \frac{{2l}}{f}} \right)\pi }}} \right.\\
 + \frac{{{{\left( { - 1} \right)}^{n - m}}\sin \left( {\frac{{2l\pi }}{f}} \right)}}{{\left( {n - m + \frac{{2l}}{f}} \right)\pi }} - \frac{{{{\left( { - 1} \right)}^{n + m + 1}}\sin \left( {\frac{{2l\pi }}{f}} \right)}}{{\left( {n + m - \frac{{2l}}{f}} \right)\pi }}\\
 + i\frac{{{{\left( { - 1} \right)}^{n + m}}\cos \left( {\frac{{2l\pi }}{f}} \right) - 1}}{{\pi \left( {n + m - \frac{{2l}}{f}} \right)}} + i\frac{{{{\left( { - 1} \right)}^{n - m}}\cos \left( {\frac{{2l\pi }}{f}} \right) - 1}}{{\pi \left( {n - m + \frac{{2l}}{f}} \right)}}\\
\left. { - i\frac{{{{\left( { - 1} \right)}^{n + m}}\cos \left( {\frac{{2l\pi }}{f}} \right) - 1}}{{\pi \left( {n + m + \frac{{2l}}{f}} \right)}} - i\frac{{{{\left( { - 1} \right)}^{n - m}}\cos \left( {\frac{{2l\pi }}{f}} \right) - 1}}{{\pi \left( {n - m - \frac{{2l}}{f}} \right)}}} \right].
\end{array}
\end{equation}
$ V_{nm}^{  d  u}$ is the conjugate of $V_{nm}^{ u  d}$.
And
\begin{equation}
\begin{array}{l}
V_{nm}^{ d  d } = \frac{{4m}}{{a{\hbar ^2}}}\int {\left[ {\sin \left( {\frac{f}{{2l}}n\bar qx} \right)\left( {\begin{array}{*{20}{c}}
0&1
\end{array}} \right)\left( {\tilde J{{\bf{n}}_r} \cdot {\bf{\sigma }} + {V_0}d} \right)\sin \left( {\frac{f}{{2l}}m\bar qx} \right)\left( {\begin{array}{*{20}{c}}
0\\
1
\end{array}} \right)} \right]} dx\\
 = \frac{{4m}}{{a{\hbar ^2}}}\int {\sin \left( {\frac{f}{{2l}}n\bar qx} \right)} {V_0}d\sin \left( {\frac{f}{{2l}}m\bar qx} \right)dx = \frac{{2m{V_0}d}}{{{\hbar ^2}}}{\delta _{n,m}}.
\end{array}
\end{equation}

\clearpage

\begin{figure}[h]
\includegraphics[height=15cm, width=9cm]{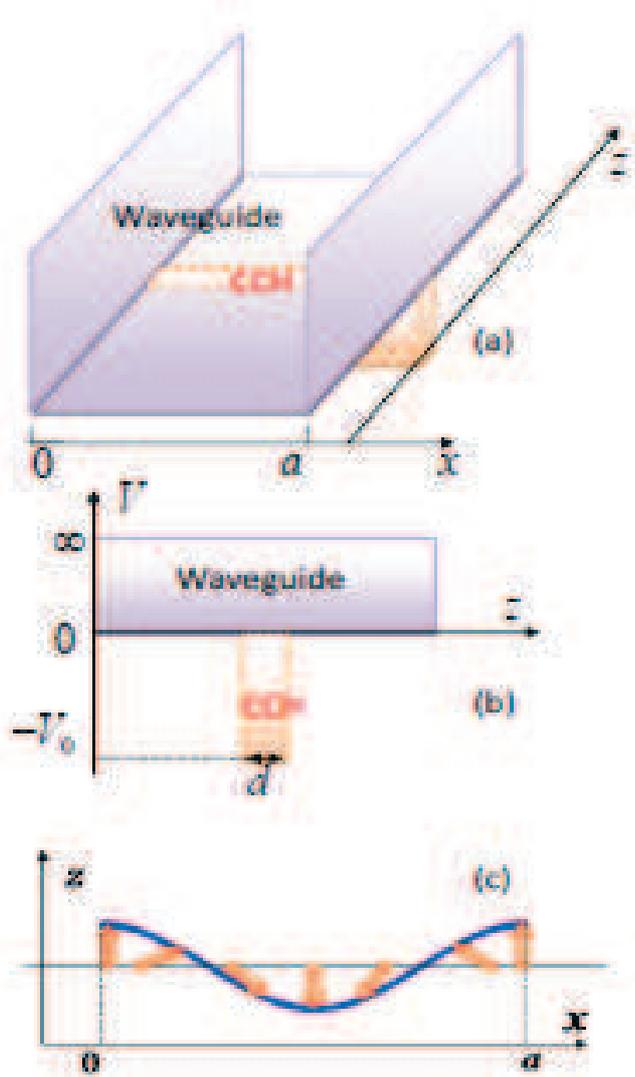}
\caption{Schematics of the proposed device. As shown in panel (a), propagation direction of the quasi-1D waveguide is in the $z$-coordinate with lateral width $a$ and infinite confining potential. Panel (b) is the side intersection of the energy profile. The conducting chiral helimagnet (CCH) layer has thickness $d$ and potential-well-depth $V_0$. The spiral spin structure of the CCH is sketched in panel (c). The spin in one atomic layer of the $x$-$y$ plane spirals in the $x$-direction with the local magnetization direction ${\bf{n}}_{\bf{r}}  = \left[ {\sin \bar qx,0,\cos \bar qx} \right]$, where $\bar q =2 \pi /a$ is the spin wave vector of the spiral.}
\end{figure}

\clearpage

\begin{figure}[h]
\includegraphics[height=13cm, width=18cm]{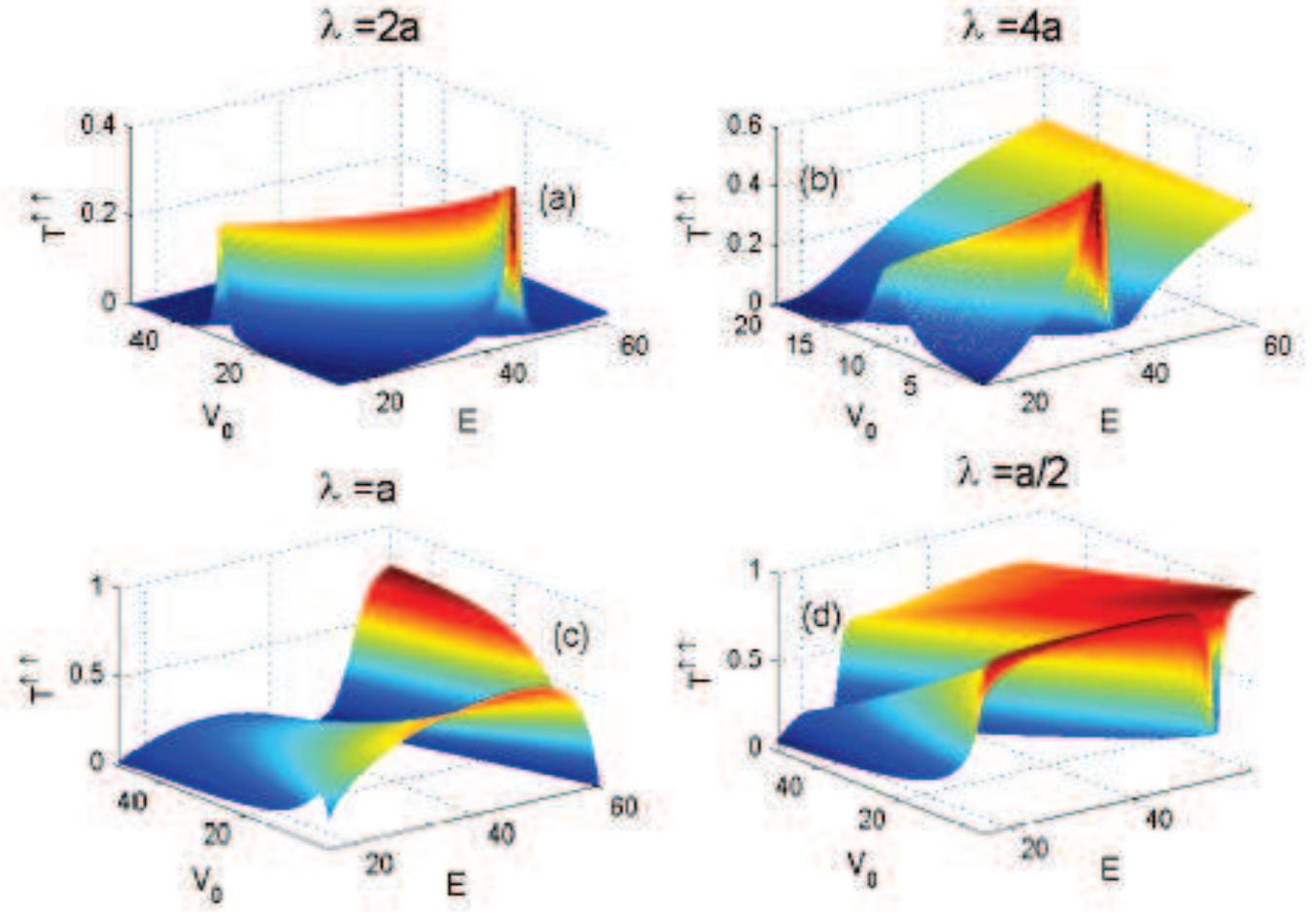}
\caption{Spin-conserved transmission $T^{u u}$ as a function of the incident energy $E$ and the CCH well depth $V_0$ in the CCH-film-embedded quasi-1D electron waveguide. The four panels are for different spiral periods $\lambda $. As discussed in the text, only the propagating subband (the first one) is considered. Here, $T^{u u}=T^{d d}$. }
\end{figure}

\begin{figure}[h]
\includegraphics[height=13cm, width=18cm]{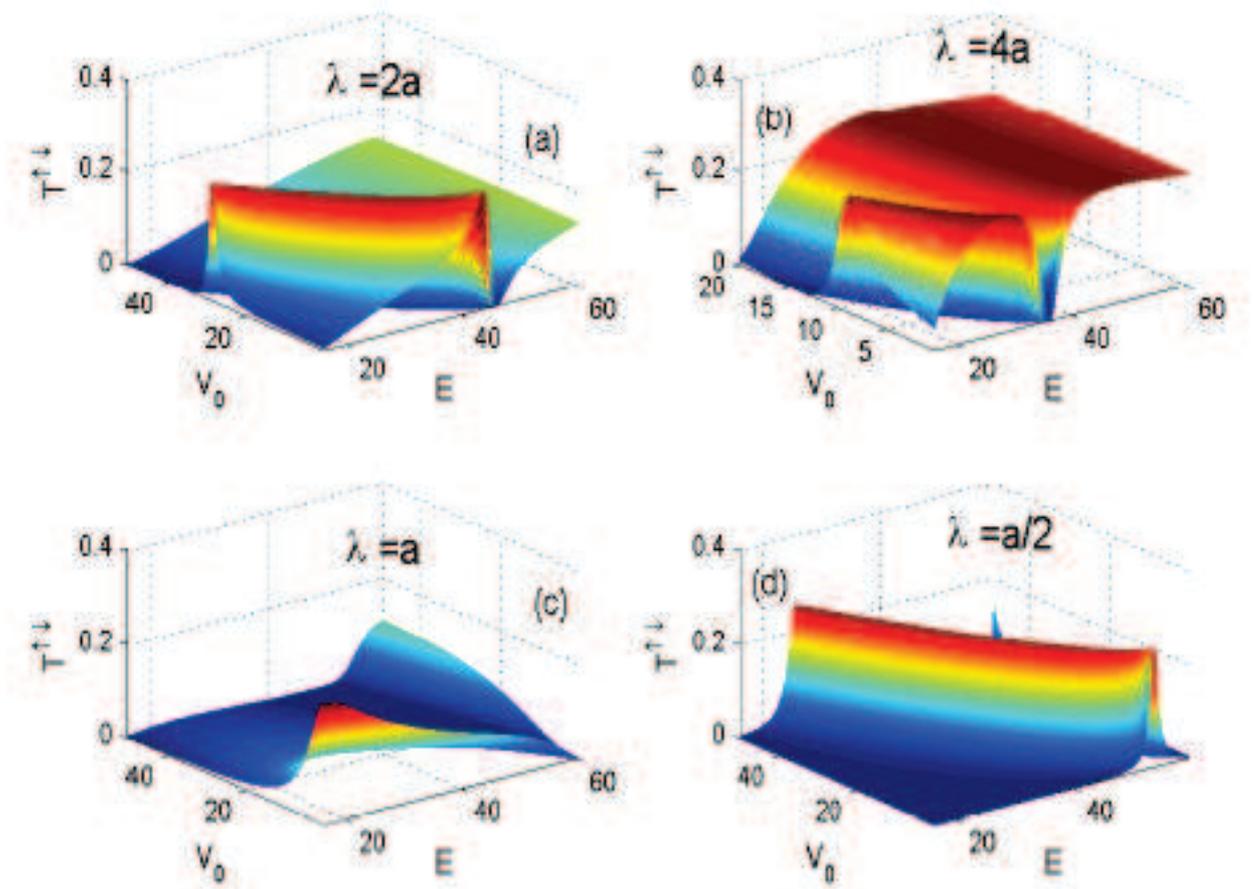}
\caption{Spin-flipped transmission $T^{u d}$ as a function of the incident energy $E$ and the CCH well depth $V_0$ in the CCH-film-embedded quasi-1D electron waveguide. As in Fig 2, the four panels are for different $\lambda $ and only the propagating subband is considered. Here, $T^{u d}=T^{d u}$. }
\end{figure}

\begin{figure}[h]
\includegraphics[height=10cm, width=14cm]{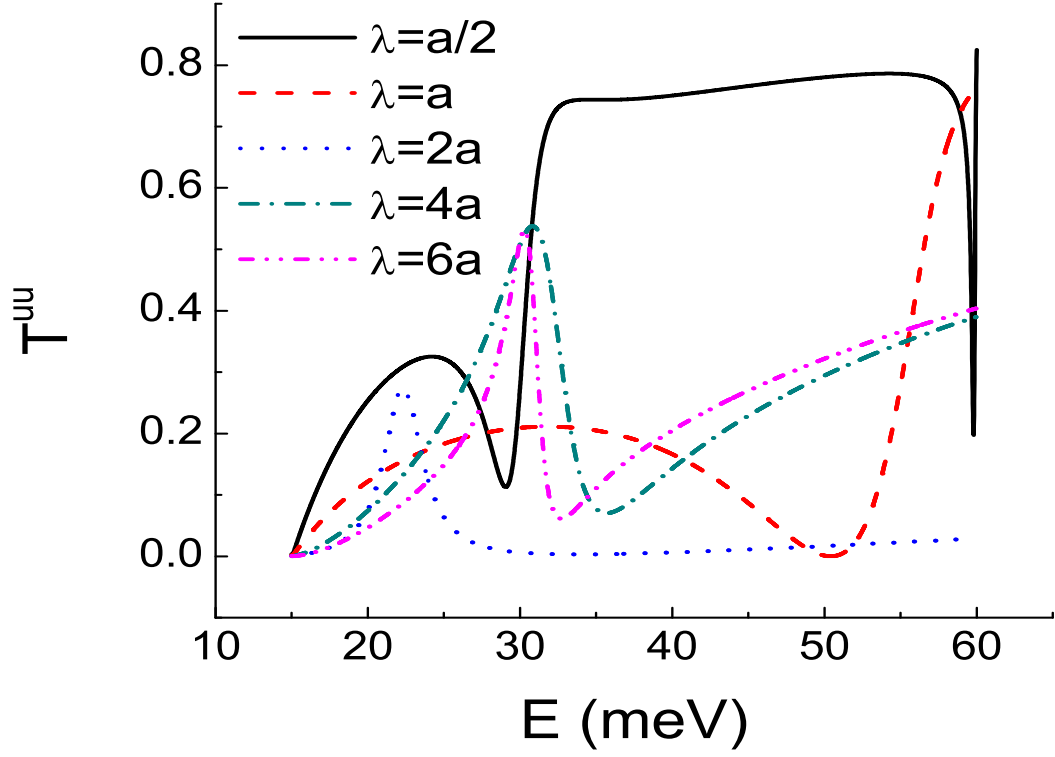}
\caption{Spin-conserved transmission $T^{u u}$ as a function of the incident energy $E$ for different $\lambda$ in the CCH-film-embedded quasi-1D electron waveguide. The CCH-well-depth $V_0 =19$ meV for $\lambda =a/2$ and $a$; $V_0 =14$ meV for $\lambda =2a$; and $V_0 =3$ meV for $\lambda =4a$ and $6a$. As in previous figures, only the propagating subband is considered. Also, $T^{u u}=T^{d d}$.}
\end{figure}

\begin{figure}[h]
\includegraphics[height=10cm, width=14cm]{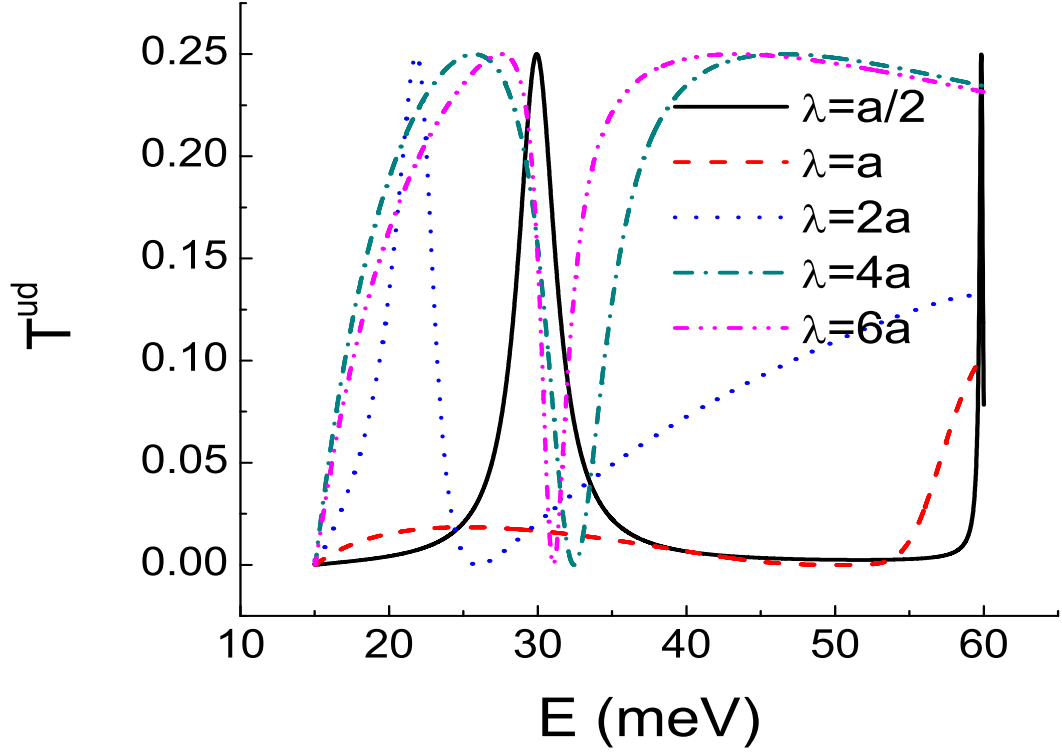}
\caption{Spin-flipped transmission $T^{u d}$ as a function of the incident energy $E$ for different $\lambda$ in the CCH-film-embedded quasi-1D electron waveguide. The CCH-well-depth $V_0$ is the same as Fig. 4. As in previous figures, only the propagating subband is considered. Also, $T^{u d}=T^{d u}$.}
\end{figure}

\begin{figure}[h]
\includegraphics[height=10cm, width=14cm]{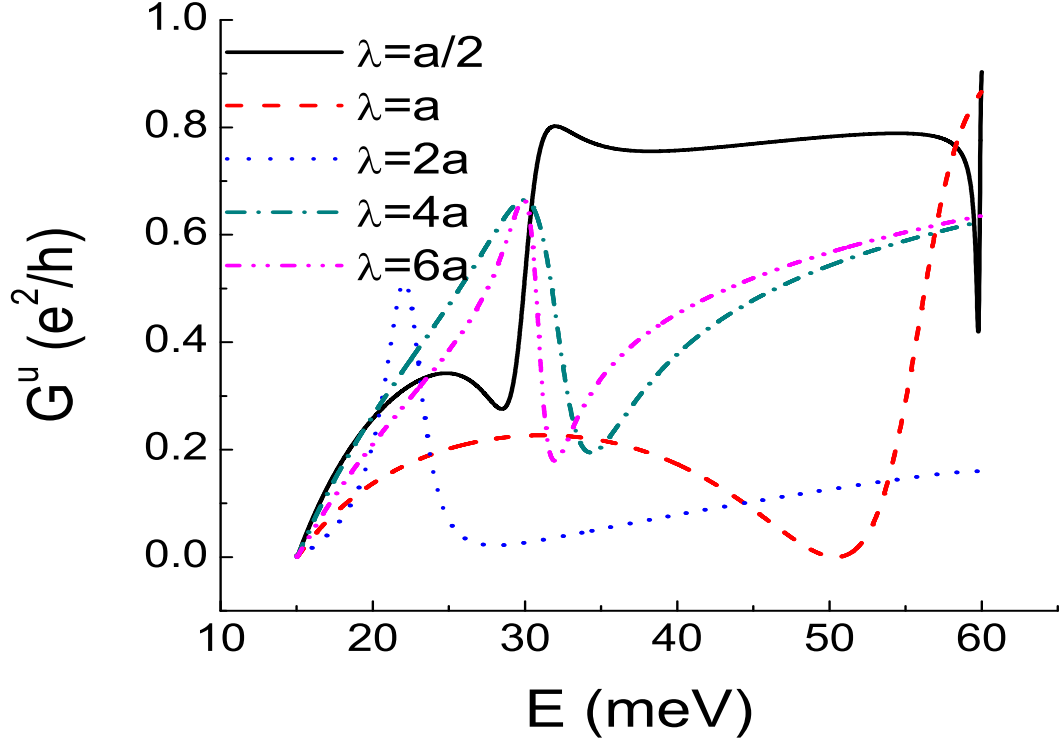}
\caption{Spin-up conductance $G^{u }$ as a function of the incident energy $E$ for different $\lambda$ in the CCH-film-embedded quasi-1D electron waveguide. Parameter settings are the same to Fig. 4. Also, $G^{u }=G^{d}$.}
\end{figure}

\begin{figure}[h]
\includegraphics[height=10cm, width=14cm]{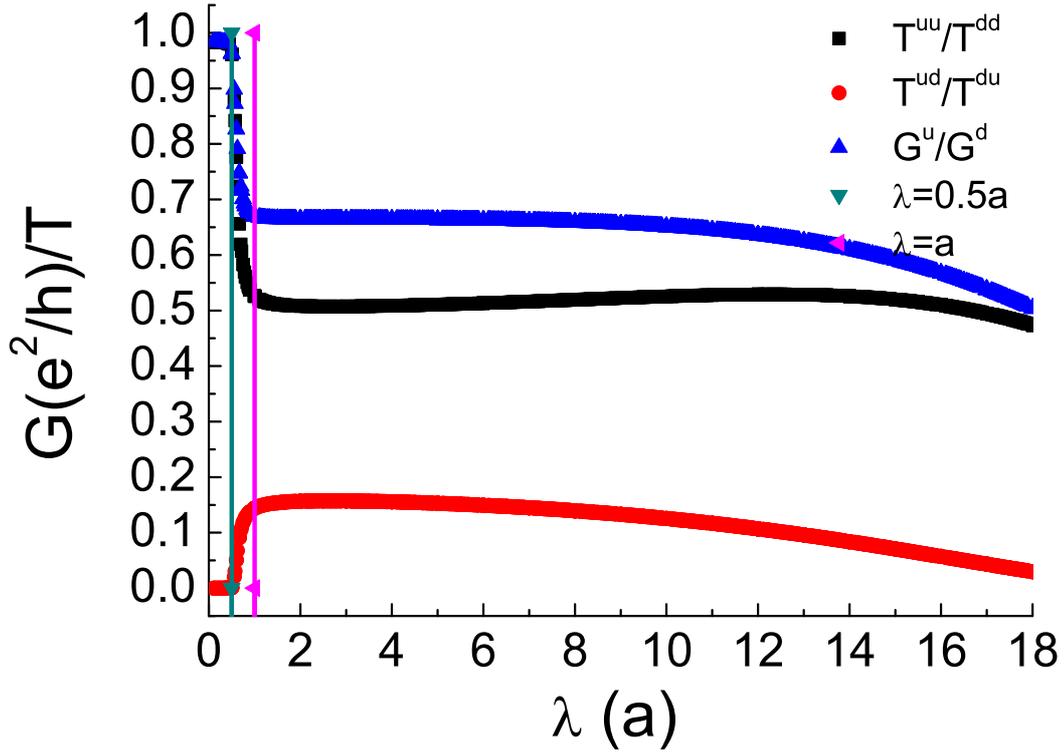}
\caption{Transmission and conductance as a function of the CCH spiral period $\lambda$ for fixed waveguide width $a=5$ nm in the CCH-film-embedded quasi-1D electron waveguide. The waveguide depth $V_0 =3$ meV and the incident electron energy $E=30$ meV. Also, $T^{uu}=T^{dd}$, $T^{ud}=T^{du}$, and $G^{u }=G^{d}$. The green and magenta lines highlight the sharp change at $\lambda = a/2$ and $a$ in the conductance.}
\end{figure}

\begin{figure}[h]
\includegraphics[height=10cm, width=14cm]{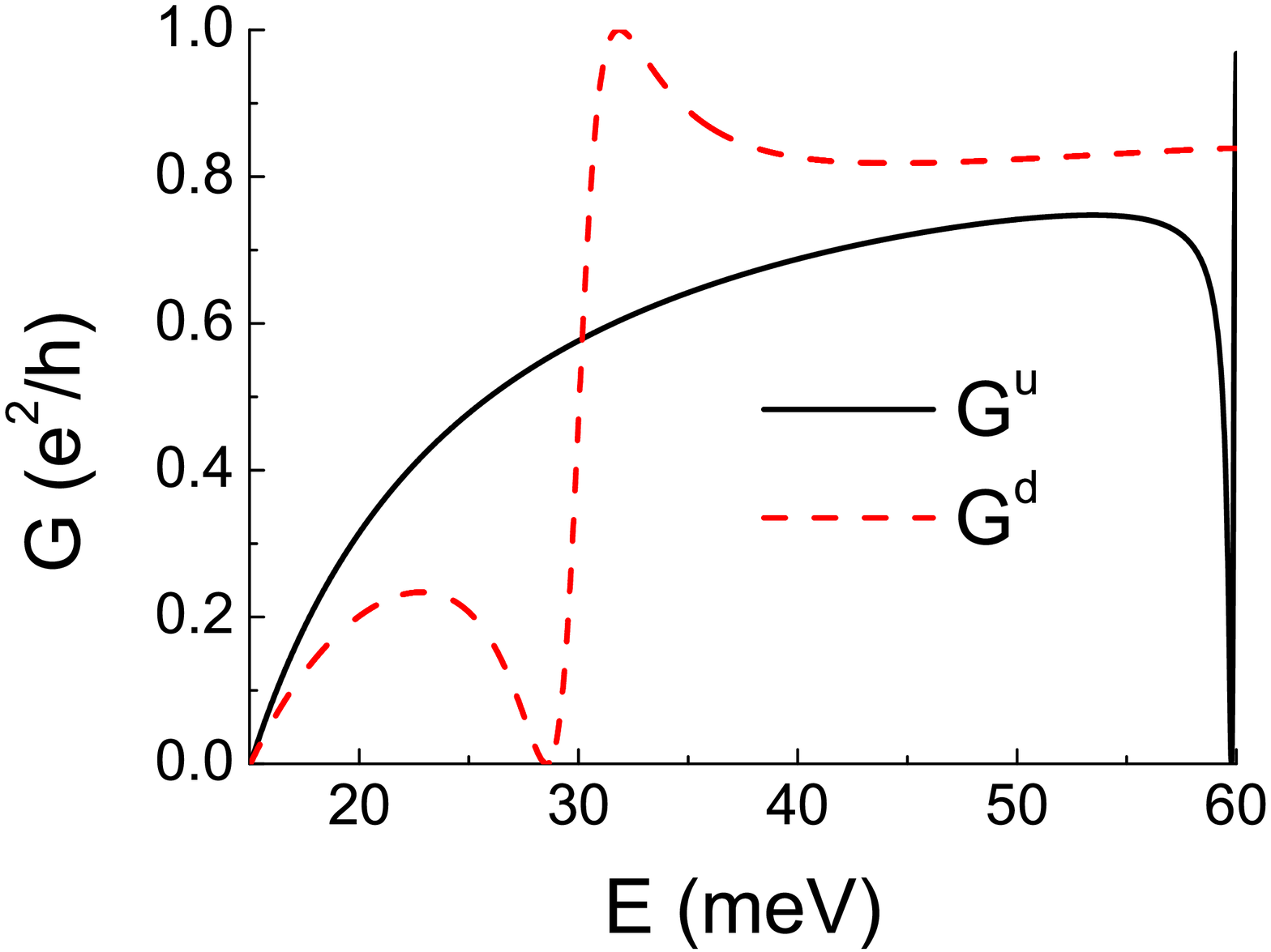}
\caption{In $\sigma _z$ representation, spin-up and down conductance $G^{u }$ and $G^{d}$ as a function of the incident energy $E$ for $\lambda =a/2$ in the CCH-film-embedded quasi-1D electron waveguide. $V_0=19$ meV.}
\end{figure}

\clearpage

\clearpage


\clearpage




\begin{references}


\bibitem{Ref1} U. Fano, Phys. Rev. \textbf{124}, B1866 (1961).

\bibitem{Ref2} E. Tekman and P. F. Bagwell, Phys. Rev. B \textbf{48}, 2553 (1993).

\bibitem{Ref8} A. E. Miroshnichenko, S. Flach, Y. S. Kivshar, Rev. Mod. Phys. \textbf{82}, 2257 (2010).

\bibitem{Ref9} H. G. Luo, T. Xiang, X. Q. Wang, Z. B. Su, and L. Yu, Phys. Rev. Lett. \textbf{92}, 256602 (2004).

\bibitem{Ref10} A. C. Johnson, C. M. Marcus, M. P. Hanson, and A. C. Gossard, Phys. Rev. Lett. \textbf{93}, 106803 (2004).

\bibitem{Ref6} B. R. Bu\l ka and P. Stefa\'{n}ski, Phys. Rev. Lett. \textbf{86}, 225128 (2001).

\bibitem{Ref23} M. E. Torio, K. Hallberg, S. Flach, A. E. Miroshnichenko, and M. Titov, Eur. Phys. J. B \textbf{37}, 399 (2004).

\bibitem{Ref3} C. S. Chu and R. S. Sorbello, Phys. Rev. B \textbf{40}, 5941 (1989).

\bibitem{Ref4} T. B.Boykin, B. Pezeshki, and J. S. Harris, Phys. Rev. B \textbf{46}, 12769 (1992).

\bibitem{Ref5} W. Porod, Z. Shao, and C. S. Lent, Appl. Phys. Lett. \textbf{61},
1350 (1992).

\bibitem{Ref24} J. L. Cardoso and P. Pereyra, Europhys. Lett. \textbf{83}, 38001 (2008).

\bibitem{Ref11} K. Kobayashi, H. Aikawa, A. Sano, S. Katsumoto, and Y. Iye,
Phys. Rev. B \textbf{70}, 035319 (2004).

\bibitem{Ref7} B. Luk¡¯yanchuk, N. I. Zheludev, S. A. maier, N. J. Halas, P. Nordlander,
H. Giessen, and C. T. Chong, Nature Materials \textbf{9}, 707 (2010).

\bibitem{Ref13} J. I. Kishine and A. S. Ovchinnikov, Phys. Rev. B \textbf{79}, 220405(R) (2009).

\bibitem{Ref19} J. Heurich, J. K\"{o}nig, and A. H. MacDonald, Phys. Rev. B \textbf{68}, 064406 (2003).

\bibitem{Ref17} C. Jia and J. Berakdar, Appl. Phys. Lett. \textbf{95}, 012105 (2009).

\bibitem{Ref18} C. Jia and J. Berakdar, Phys. Rev. B \textbf{81}, 052406 (2010).

\bibitem{Ref14} A. Manchon, A. Pertsova, N. Ryzhanova, A. Vedyayev, and
B. Dieny, J. Phys.: Condens. Matter \textbf{20}, 505213 (2008).

\bibitem{Ref15} A. Manchon, N. Ryzhanova, A. Vedyayev, and B. Dieny, J. Appl. Phys. \textbf{103}, 07A721 (2008);

\bibitem{Ref16} R. Zhu, arXiv:1204.6095.

\bibitem{Ref12} J. I. Kishine, I. V. Proskurin, and A. S. Ovchinnikov, Phys. Rev. Lett. \textbf{107}, 017205 (2011).

\bibitem{Ref25} Here actually $V_0 d$ is the $\delta$-barrier strength.
If compared to a finite-width barrier, it is a $V_0$-high and $d$-wide barrier. We tuned the strength of $V_0 d$ in numerical treatment.

\bibitem{Ref20} P. F. Bagwell, Phys. Rev. B \textbf{41}, 10354 (1990).

\bibitem{Ref21} N. Kanazawa, Y. Onose, T. Arima,
D. Okuyama, K. Ohoyama, S. Wakimoto, K. Kakurai, S. Ishiwata, and Y.
Tokura, Phys. Rev. Lett. \textbf{106}, 156603 (2011).

\bibitem{Ref22} K. A. Bachman, J. J. Peltzer, P. D. Flammer, T. E. Furtak, R. T. Collins, and R. E. Hollingsworth, Optics Express \textbf{20}, 1308 (2012).


\end{references}
\end{document}